# PRESENT STATUS OF EPSILON′/EPSILON

Mohammad Saleem, Haris Rashid, Faisal Akram and Atif Shahbaz

Centre for High Energy Physics, Punjab University, Lahore. PAKISTAN.

## ABSTRACT

We give the recent experimental results and theoretical analysis of CP violating ratio epsilon′/epsilon. The present status is discussed in the light of the standard model.

The first observation of CP violation in the decay of neutral kaons was reported in 1964 [1]. This occurred via the mixing of CP eigenstates and was called indirect CP violation. The observation of direct CP violation which takes place in the decay process itself was very difficult to measure. This is described by the parameter Re $\varepsilon′/\varepsilon$ where $\varepsilon′$ projects the direct CP violation while $\varepsilon$ is associated with the indirect CP violation. The first experimental result showing the existence of direct CP violation was obtained in 1988 by Burkhardt et al.[2]. The experiment was performed by studying the decay of kaons which come in two types: long-lived kaons, $K_L$, and short-lived kaons, $K_S$. The final results published by NA31 and by E731 Collaborations are [3]

$$\text{Re } \varepsilon′/\varepsilon = (23.0 \text{ " } 6.5) \times 10^{-4}$$

$$\text{Re } \varepsilon′/\varepsilon = (7.4 \text{ " } 5.9) \times 10^{-4},$$

respectively. The former result, although a very small number, is definitely different from zero and therefore exhibits the direct CP violation while the latter result could be compatible with no direct CP violation. It was therefore natural to design experiments which would give more precise results. NA48 at CERN and KTeV at Fermilab were such experiments. The most recent values obtained in these experiments are [4]:

$$\text{Re } \varepsilon′/\varepsilon = (15.3 \text{ " } 2.6) \times 10^{-4}$$

$$\text{Re } \varepsilon′/\varepsilon = (20.7 \text{ " } 2.8) \times 10^{-4},$$

respectively. The result at NA48-CERN was computed with 3.3 million $K \rightarrow \pi^0 \pi^0$ events collected during the 1998 and 1999 running periods. The last two results were obtained by NA48 and KTeV using the same experimental principle. That is in each experiment the double ratio of $K_L$, and $K_S$ going to neutral particles and to charged particles was measured. However, there are significant differences in the experimental and analysis techniques of the two experiments. The new world average value of Re $\varepsilon′/\varepsilon$ is $(17.2 \text{ " } 1.8) \times 10^{-4}$ [5]. This confirms the existence of direct CP violation in the neutral kaon system. It may be noted that the $\varepsilon′/\varepsilon$ programme of NA48 has been completed with the successful 2001 data-taking [6]. The new data will improve the statistical accuracy and perform a major check of its present result. Let us now see what is the theoretical status of $\varepsilon′/\varepsilon$.

The first estimate of $\varepsilon′/\varepsilon$ was made more than a quarter of a century ago. It was only by chance that although calculations were made on simple basis ignoring some important effects the result was about 1/450, a value compatible with the present world average. Significant progress has been made since then. Hambye et al [7] presented an analysis of the ratio $\varepsilon'/\varepsilon$. They calculated the hadronic matrix elements using the $1/N_C$ expansion within the framework of the effective chiral Lagrangian for pseudoscalar meson at leading plus next-to- leading order in the chiral expansion. They obtained the results for three sets of Wilson coefficients. The result for the set LO, for $\Lambda_C$ between 600 and 900 MeV and for central values of the parameters is

$$14.8 \times 10^{-4} \# \varepsilon'/\varepsilon \# 19.4 \times 10^{-4}.$$

.

This technique was improved by Pallante, Scimemi and Pich [8] who in 2001 presented a detailed analysis of $\varepsilon'/\varepsilon$ within the standard model, taking into account final state interactions. A matching procedure between the effective short-distance Lagrangian and its corresponding low-energy description in chiral perturbation theory was used to fix the relevant hadronic matrix elements at leading order in the $1/N_C$ expansion. They obtained

$$\text{Re } \varepsilon'/\varepsilon = (17" \ 9) \times 10^{-4}.$$

The error is dominated by the uncertainty in the mass of the strange quark. A better estimate of this mass would reduce the uncertainty up to about 30%. It may be emphasised that the mass of a strange quark is crucial for the precise computation of $\varepsilon'/\varepsilon$. Moreover, the calculations of hadronic matrix elements have yet to reach a level of the NLO calculations of the Wilson coefficients. The uncertainty in the final theoretical result arises due to the uncertainty in the calculations of these quantities.

Wu [9] has made a prediction about $\varepsilon'/\varepsilon$ by investigating the low energy dynamics of QCD with special attention to the matching between QCD and chiral perturbation theory, and also to some useful algebraic chiral operator relations which survive even when the chiral loop corrections are included. A value of

$$\varepsilon'/\varepsilon = (20 \ 4 \ " \ 5 \ " \ 4) \times 10^{-4}$$

was obtained. This is also consistent with the world average.

Buras and Gerard [5] have shown that the recently obtained experimental value for $\varepsilon'/\varepsilon$ does not require sizable $1/N_C$ and isospin-breaking corrections. For long-distance hadron physics, isospin symmetry and large NC limit are two powerful approximations. By making use of them, they have shown that

$$\text{Re } \varepsilon'/\varepsilon = (17.4 \ " \ 0.7) \times 10^{-4}.$$

## Acknowledgment

Two of us (Haris Rashid and Atif Shahbaz) are grateful to the Pakistan Science Foundation for financial assistance under Project No. P-PU/PHYS (117).